\documentclass[aps,prl,preprint, superscriptaddress, onecolumn]{revtex4-2}
\usepackage[utf8]{inputenc}
\usepackage{hyperref}
\usepackage{graphicx}
\usepackage{xcolor}
\usepackage{natbib}

\usepackage{amsfonts, amsmath, bm, graphicx}
\usepackage{siunitx}
\DeclareSIUnit\angstrom{\text {Å}}
\let\DeclareUSUnit\DeclareSIUnit

\DeclareUSUnit\dBm{dBm}
\usepackage{float}
\usepackage{comment}

\usepackage{color,soul}
\usepackage{textgreek}
\usepackage{textcomp}
\usepackage{gensymb}

\begin{document}

\author{Atul Pandey}
\affiliation{Max Planck Institute of Microstructure Physics, Weinberg 2, 06120 Halle, Germany}
\affiliation{Institute of Physics, Martin Luther University Halle-Wittenberg, Von-Danckelmann-Platz 3, 06120 Halle, Germany} 
\author{Jitul Deka}
\affiliation{Max Planck Institute of Microstructure Physics, Weinberg 2, 06120 Halle, Germany}
\author{Jiho Yoon}
\affiliation{Max Planck Institute of Microstructure Physics, Weinberg 2, 06120 Halle, Germany}
\author{Chris Koerner}
\affiliation{Institute of Physics, Martin Luther University Halle-Wittenberg, Von-Danckelmann-Platz 3, 06120 Halle, Germany} 
\author{Rouven Dreyer}
\affiliation{Institute of Physics, Martin Luther University Halle-Wittenberg, Von-Danckelmann-Platz 3, 06120 Halle, Germany}
\author{James M. Taylor}
\affiliation{Institute of Physics, Martin Luther University Halle-Wittenberg, Von-Danckelmann-Platz 3, 06120 Halle, Germany}
\author{Stuart S. P. Parkin}
\affiliation{Max Planck Institute of Microstructure Physics, Weinberg 2, 06120 Halle, Germany}
\author{Georg Woltersdorf}
\affiliation{Institute of Physics, Martin Luther University Halle-Wittenberg, Von-Danckelmann-Platz 3, 06120 Halle, Germany} 
\affiliation{Max Planck Institute of Microstructure Physics, Weinberg 2, 06120 Halle, Germany}
\email{georg.woltersdorf@physik.uni-halle.de}
\date{\today}
\title{Anomalous Nernst effect based near field imaging of magnetic nanostructures}

\begin{abstract}
The anomalous Nernst effect (ANE) gives rise to an electrical response transverse to the magnetization and an applied temperature gradient in a magnetic metal. A nanoscale temperature gradient can be generated by the use of a laser beam applied to the apex of an atomic force microscope tip, thereby allowing for spatially-resolved ANE measurements beyond the optical diffraction limit. Such a method has been used previously to map in-plane magnetized magnetic textures. However, the spatial distribution of the out-of-plane temperature gradient, which is needed to fully interpret such an ANE-based imaging, was not studied. We therefore use a well-known magnetic texture, a magnetic vortex core, to demonstrate the reliability of the ANE method for the imaging of magnetic domains with nanoscale resolution. Moreover, since the ANE signal is directly proportional to the temperature gradient, we can also consider the inverse problem and deduce information about the nanoscale temperature distribution. Our results together with finite element modeling indicate that besides the out-of-plane temperature gradients, there are even larger in-plane temperature gradients. Thus we extend the ANE imaging to study out-of-plane magnetization in a racetrack nano-wire by detecting the ANE signal generated by in-plane temperature gradients. In all cases, a spatial resolution of about 80 nm is obtained. These results are significant for the rapidly growing field of thermo-electric imaging of antiferromagnetic spintronic device structures.    
\end{abstract}
\maketitle

The miniaturization of spintronic devices requires individual magnetic entities to be densely packed. Magnetic stray fields due to interactions between neighboring bits are a major limitation for the packing density when nano-scale ferromagnets are used \cite{Packing_Hu2011,AS_rev}. An elegant solution that has been used since the very first spintronic sensors and in magnetic random access memories is a synthetic antiferromagnet \cite{SSPP_SAF} formed from thin ferromagnetic layers coupled via a thin metallic antiferromagnetic coupling layer. Recently, there has been increased interest in utilizing innately antiferromagnetic (AF) materials \cite{AS_rev,NCAFM_nat1,AFM_MTJ_Chen2023,AFM_MTJ_Shao2024}, which are free from stray fields. The absence of stray fields makes it very difficult to both calculate the size and to image AF domains.  Typically, it is understood that these materials exhibit domain structures at the sub-micrometer scale \cite{AFM_dom_size,AFM_dom_kerr_Higo2018}. In order to understand their behavior (e.g. their response to magnetic fields and spin torques or interaction with structural defects) imaging of the magnetic order with nanometer resolution is required. Magneto-optical imaging techniques are limited in their resolution to the wavelength of the light \cite{MOKE_Soldatov2017}, and X-ray based photoemission microscopy \cite{PEEM_science_CuMnAs_jungwirth,PEEM_Mn2Au,PEEM_Mn2Au_PRB,PEEM_Reimers2023} is extremely surface sensitive.  

The magnetic state of both ferromagnetic and antiferromagnetic materials can be inferred from their off-diagonal magnetotransport behaviors \cite{NCAFM_nat2,Atul_PRB,AFM_Pal2022}. One important example is the anomalous Nernst effect (ANE) that generates an electric field (\textbf{E}) transverse to both the magnetization (\textbf{M}) and an applied temperature gradient ($\mathrm{\mathbf{\nabla}T}$) \cite{ANE}. By using laser heating to generate a local $\mathrm{\mathbf{\nabla}T}$ that can be rastered across the sample, spatially-resolved measurements of ANE-generated voltage ($V_{\mathrm{ANE}}$) allow for the imaging of magnetic domains. This method of scanning ANE (SANE) microscopy has previously been used to image the magnetic domains exhibited by in-plane (IP) magnetized thin-film devices with a spatial resolution of a few micrometers \cite{ANE_Mic,ANE_mic1,ANE_mic2}. More recently, the spatial resolution has been improved by using a nanoscale metallic tip as a near-field antenna\cite{ANE_mic_SNOM,ANE_mic_SNOM1,ANE_mic_SNOM2}. In practice, a metallized tip of an atomic force microscope (AFM) is used to confine the laser heating to a nanoscale region under the tip.
 
Here, we build on previous work to extend ANE-based near-field microscopy (NF-SANE) to out-of-plane magnetization imaging. This is inspired by finite element simulations of the laser heating on the nanoscale. Surprisingly, we find that so far ignored in-plane (IP) component of the temperature gradient is actually twice as large as its out-of-plane (OOP) component. Based on this result, we first image the well-known in-plane magnetized Landau pattern of a magnetic vortex state by applying an OOP temperature gradient. This allows us to demonstrate the validity of the ANE microscope principle and compute the spatial distribution of the IP temperature gradient. In the second step, the magnetic domains in an OOP magnetized racetrack nanowire are imaged by using an IP temperature gradient. In both cases, the spatial resolution we obtain is $\approx 80$ nanometers. 

\begin{figure}[H]
\centering
\includegraphics[width=1\textwidth]{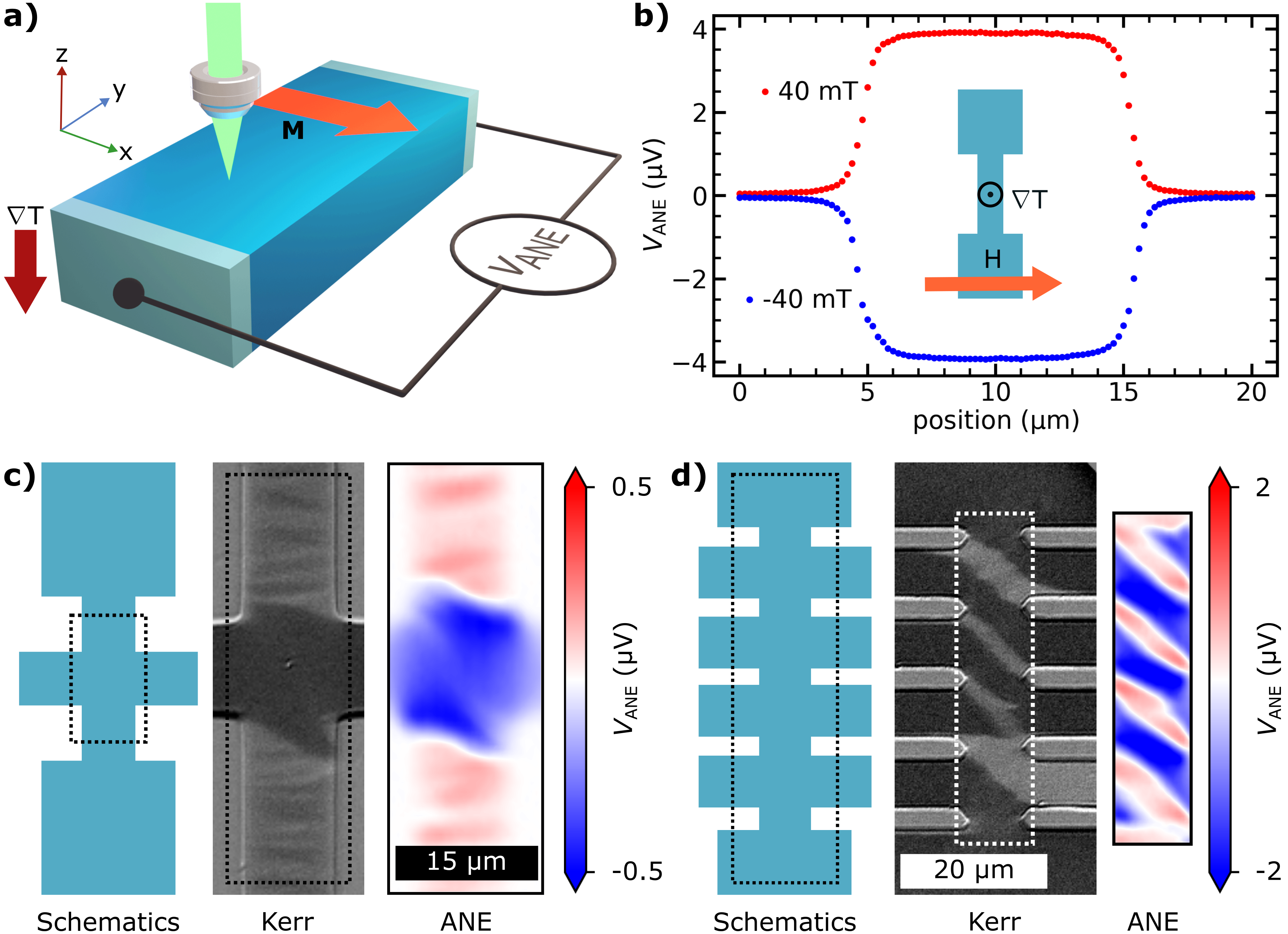}
\caption{(a) Schematic illustration of the ANE imaging method. The $V_{\mathrm{ANE}}$ is given by the transverse component of the magnetization shown by the orange arrow and the vertical temperature gradient shown by the red arrow. (b) Line scan of $V_{\mathrm{ANE}}$ across a device of width w = \SI{10}{\micro m} in a field of \SI{+40}{mT} and \SI{-40}{mT}. The device is illuminated with a \SI{5}{mW} laser beam focused by a 60X ($NA$ = 0.7) objective. The inset shows a schematic of the device structure utilized for the measurements. (c,d) Kerr and SANE microscopy images of multi domain states stabilized in different device structures as indicated with the schematics.} 
 \label{fig:ff_method}
\end{figure}

The electrical response resulting from the ANE is given by ${\mathbf{E_{\mathrm{ANE}}}} = \mathrm{\mu_0} S_{\mathrm{ANE}}\mathbf{\nabla} T \times \mathbf{M}$, where $S_\mathrm{{ANE}}$ is the ANE coefficient. The resulting voltage measured along the device length (y-direction) is described by the following equation:
\begin{equation}\label{eq1}
    {V_{\mathrm{ANE}}} \propto \mu_0\cdot S_{\mathrm{ANE}}\cdot \nabla T_{\mathrm{z}}\cdot M_{\mathrm{x}}+\mu_0\cdot S_{\mathrm{ANE}}\cdot\nabla T_{\mathrm{x}}\cdot M_{\mathrm{z}}
\end{equation}
In order to validate the applicability of the ANE microscopy to image magnetic textures we verify the magnetic origin of the laser-induced signals and compare magneto-optic microscopy and SANE microscopy \cite{ANE_Mic,ANE_mic1,ANE_mic2} images of magnetic microstructures with domains.
Here we use a laser beam focused by a microscope objective to create a temperature gradient while scanning the sample laterally for imaging. A \SI{15}{nm} thick IP magnetized $\mathrm{Co_{20}Fe_{60}B_{20}}$ film is patterned into a \SI{10}{\micro m} wide wire as shown with the schematic \textbf{Fig.\ref{fig:ff_method}(b)}. A magnetic field $\mathrm{\mu_0 H}$ is applied along the width of the wire. We observe a $V_{\mathrm{ANE}}$ signal on the order of a few $\mathrm{\micro V}$ as the laser beam is scanned across the wire. This signal clearly is of magnetic origin as its polarity changes when magnetic field is reversed in direction.

Next, we show that spatially-resolved ANE measurements can be utilized to image the magnetic domain structure of a sample. For this, we use similar \SI{15}{nm} thick CoFeB wire structures with branches, as shown in \textbf{Fig.\ref{fig:ff_method}(c,d)}. These branches stabilize a multi-domain remnant state, as confirmed by static Kerr microscopy. The corresponding ANE measurements well reproduce these domain structures with a much better contrast.  

\begin{figure}[H]
\centering
\includegraphics[width=1\textwidth]{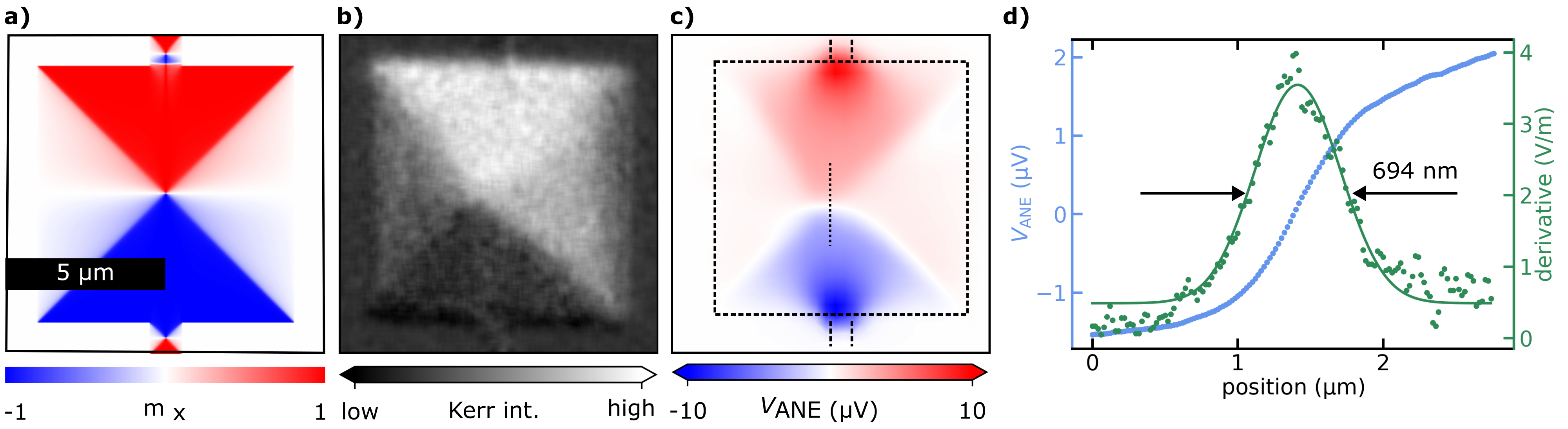}
\caption{(a) Micromagnetic simulation of an $8 \times 8$ $\mathrm{\micro m^2}$ CoFeB slab of \SI{45}{nm} thickness fashioned with \SI{1}{\micro m} wide contact electrodes at top and bottom. (b) Longitudinal Kerr intensity image showing the $x$ component of the magnetisation of the film. (c) ANE microscope image of the magnetic domains in the same device as (b), the outline of which is shown by the dashed box. (d) Line scan of the ANE signal across the center of the vortex, indicated by the dotted line in (c), right (green) axis shows derivative of the line scan fitted to a Gaussian distribution.} 
\label{fig:ff_domain}
\end{figure}

Applying this method to a nanoscale spin texture of a magnetic vortex \cite{Magnetic_vortex_nature_2006,Magnetic_vortext_science_2002,Magnetic_vortex_science2000} allows us to understand the spatial spreading of the thermal heat gradients. For this, we use a $8 \times 8$ $\mathrm{\micro m^2}$ CoFeB slab of \SI{45}{nm} thickness, with \SI{1}{\micro m} wide contact wires (top and bottom in the image in \textbf{Fig.\ref{fig:ff_domain}}) . The ANE signal, along the y-direction that is proportional to $\mathrm{m_x}$, is measured using these contacts.
A differential Kerr microscope image, shown in \textbf{Fig.\ref{fig:ff_domain}(b)}, reveals the expected  pattern around the vortex core this structure. SANE imaging well reproduces this pattern, as demonstrated in \textbf{Fig.\ref{fig:ff_domain}(c)}. We note that the ANE signal is higher near the wire connections. This is due to the lateral spreading of the current generated by $\mathrm{V_{ANE}}$ throughout the slab. 

A magnetic vortex structure leads to a very rapid rotation of the IP magnetization as the vortex core has a width of only a few nm\cite{Magnetic_vortext_science_2002,Magnetic_vortex_nature_2006,Magnetic_vortex_science2000}. This results in a nanoscale transition between opposing in-plane magnetization directions across the vortex. An ANE line scan through the vortex allows us to compute the spatial distribution of the heat gradient $\nabla T_{\mathrm{z}}$ (Suppl. \textbf{S.IX}), which is given by the derivative of the ANE line scan. This is well described by a Gaussian distribution with the full width at half maximum (FWHM) of $694\pm18$ nm (\textbf{Fig.\ref{fig:ff_domain}(d)}). This gradient distribution is very close to the intensity distribution of the focused laser beam used for SANE, which is also a Gaussian with a FWHM = $748\pm22$ nm (\textbf{Fig.S7(b)}). A Finite-element modelling \cite{COMSOL} analysis reveals that the OOP temperature gradient closely follows the radial distribution of the laser intensity (\textbf{Fig.S5(d)}). This is in line with our findings.

Next, we test if this is still the case on the nanoscale, i.e. when optical near-field focusing near the apex of a metallic AFM tip \cite{SNOM,SNOM1} confines the heat source to \SI{\approx 20}{nm}. Our simulations show that even for a \SI{20}{nm} wide Gaussian heat source the broadening of the heat gradient due to thermal spreading in the sample only amounts to $\approx$26\%, as shown in (\textbf{Fig.S5(d)}). This implies that heat spreading only has a small effect on the spatial resolution even for NF-SANE imaging.  

\begin{figure}[H]
\centering
\includegraphics[width=1\textwidth]{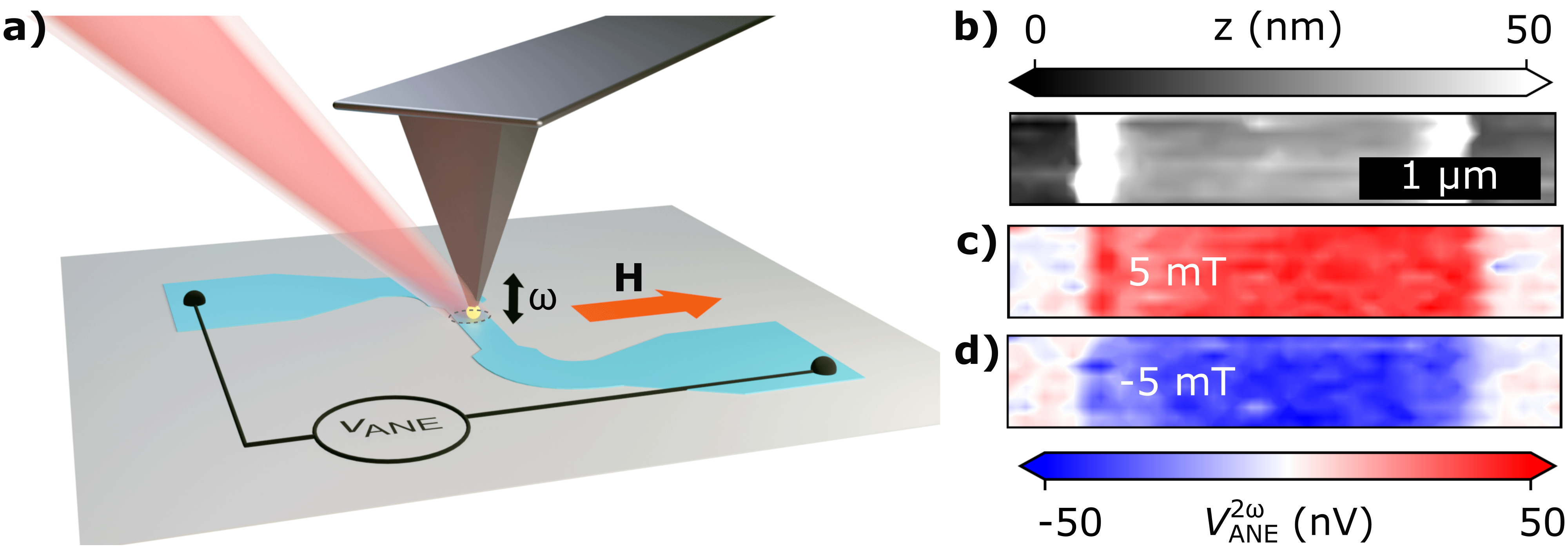}
\caption{(a) Schematic illustration of of the NF-SANE microscope. (b) AFM topographic scan of a \SI{2}{\micro m} wide magnetic wire. (c,d) ANE $\mathrm{2^{nd}}$ harmonic signal under the application of a transverse \SI{5}{mT} and \SI{-5}{mT} magnetic field, respectively.} 
\label{fig:nf_method}
\end{figure}

Thus, we now experimentally study ANE imaging with nanoscale gradients created by the enhanced optical near-field \cite{SNOM_enhancement,SNOM_enhancement1} of a metallic AFM tip (\textbf{Fig.\ref{fig:nf_method}(a)}). The tapping of the AFM tip generates an intensity-modulated optical near field that, in turn, generates an oscillating thermal gradient with a radial distribution on the nanometer scale. The enhanced optical near field at the sample has a non-linear dependence on the tip-to-sample distance. Therefore, the sinusoidal oscillation of the AFM tip results in the generation of higher harmonics of the tips oscillation frequency in the ANE signal. We record the voltage demodulated at the second harmonic of the tip oscillation frequency in order to detect the ANE signal that exclusively originates from the optical near field. In the experiment, we again use a \SI{15}{nm} thick CoFeB film now patterned into a narrower  \SI{2}{\micro m} wide wire, as shown by the AFM image in \textbf{Fig.\ref{fig:nf_method}(b)}. A magnetic field of $\pm$5 mT is applied along the $x$-direction of the wire structure. The resulting NF-SANE images for both field directions are shown in \textbf{Figs.\ref{fig:nf_method}(c, d)}, respectively. When the tip approaches the magnetic wire, a ANE signal on the order of 40 nV is observed. Again, the voltage polarity reverses with field direction, confirming the magnetic origin of the near-field signal.

\begin{figure}[H]
\centering
\includegraphics[width=1\textwidth]{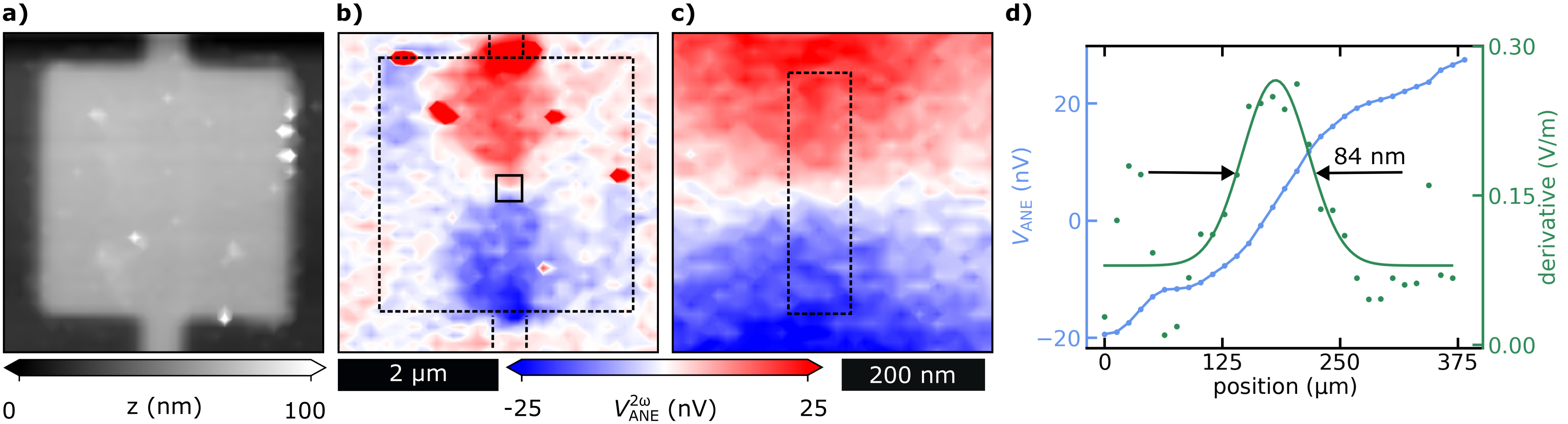}
\caption{(a) AFM height scan of a $3\times3$ $\mathrm{\micro m^2}$ square device. (b) Simultaneously measured $\mathrm{2^{nd}}$-harmonic ANE voltage. Outer dashed square indicates the edge of the device. (c) High-resolution ANE scan in an area around the magnetic vortex shown by the inner solid square in (b). (d) Average ANE signal from eight line scans measured in the dotted rectangle region shown in (c), right (green) axis shows the derivative of the line scan, fitted by a Gaussian.} 
\label{fig:nf_domain}
\end{figure}

To study the nanoscale heat gradient, we again use a Landau pattern, but on a smaller device consisting of a $3\times3$ $\mathrm{\micro m}^2$ square connected to \SI{500}{nm} wide contact wires. An AFM height scan with \SI{100}{nm} step size shows the overall structure of the device in \textbf{Fig.\ref{fig:nf_domain}(a)}. The simultaneously measured ANE signal in \textbf{Fig.\ref{fig:nf_domain}(b)} reveals a Landau pattern, representing the magnetic domain structure. A scan with a smaller step size (\SI{12}{nm}) shows the vortex core region with higher resolution (\textbf{Fig.\ref{fig:nf_domain}(c)}). 
In Fig \textbf{Fig.\ref{fig:nf_domain}(d)} we follow the same procedure as in \textbf{Fig.\ref{fig:ff_domain}(d)} and show the corresponding heat gradient distribution for the NF-SANE case. Again, we find a Gaussian distribution with a FWHM of $\mathrm{84 \pm 14}$ nm. In such a scanning measurement, the resolution is determined by the FWHM of the probe. i.e. we can expect a resolution of \SI{\approx 84}{nm} for NF-SANE. 


Until our work here, SANE or NF-SANE imaging were not used to investigate OOP magnetized structures because it was thought that a large $\nabla T$ only occurs OOP. However, our finite element simulations of the heat distribution shows that the IP temperature gradient ($\nabla T_{\mathrm{x}}$) is actually almost a factor of two larger than the OOP temperature gradient (\textbf{Fig.S6}). Considering equation (1), it is rather surprising that the IP heat gradient has not been considered to date \cite{ANE_mic_SNOM2}. Thus, in the following we analyze its significance and show that it can be used to visualize OOP magnetization on the nanoscale.

The near-field focus of light leads to a spatially non-uniform IP temperature gradient $\nabla T_{\mathrm{x}}$ (\textbf{Fig.S6}). In this case, the second term in equation \eqref{eq1} can be written as follows (Suppl.(\textbf{S.VIII})): 
\begin{equation}\label{eq2}
    V_{\mathrm{ANE}} = \frac{\mu_{0}\cdot S_{\mathrm{ANE}}\cdot l_\mathrm{spot}}{w}\int_{0}^{w}m_{\mathrm{z}}\nabla T_{\mathrm{x}} \, dx
\end{equation}
where $l$ and $w$ correspond to the length and width of the wire, respectively, and $l_{\mathrm{spot}}$ is the spot size of the heat gradient. For a uniform OOP magnetization under a heat source, equation\eqref{eq2} can be simplified as follows: 
\begin{equation}\label{eq3}
    V_{\mathrm{ANE}} = \mu_{0} \cdot  S_{\mathrm{ANE}}\cdot l_\mathrm{spot} \cdot m_{\mathrm{z}}\cdot \nabla T_{\mathrm{x}}^{\mathrm{avg}}
\end{equation}
Where $\nabla T_{\mathrm{x}}^{\mathrm{avg}}=\int_0^w\frac{\nabla T_{\mathrm{x}}}{w}\, dx$ is the average IP temperature gradient in the wire. 

\begin{figure}[H]
\centering
\includegraphics[width=1\textwidth]{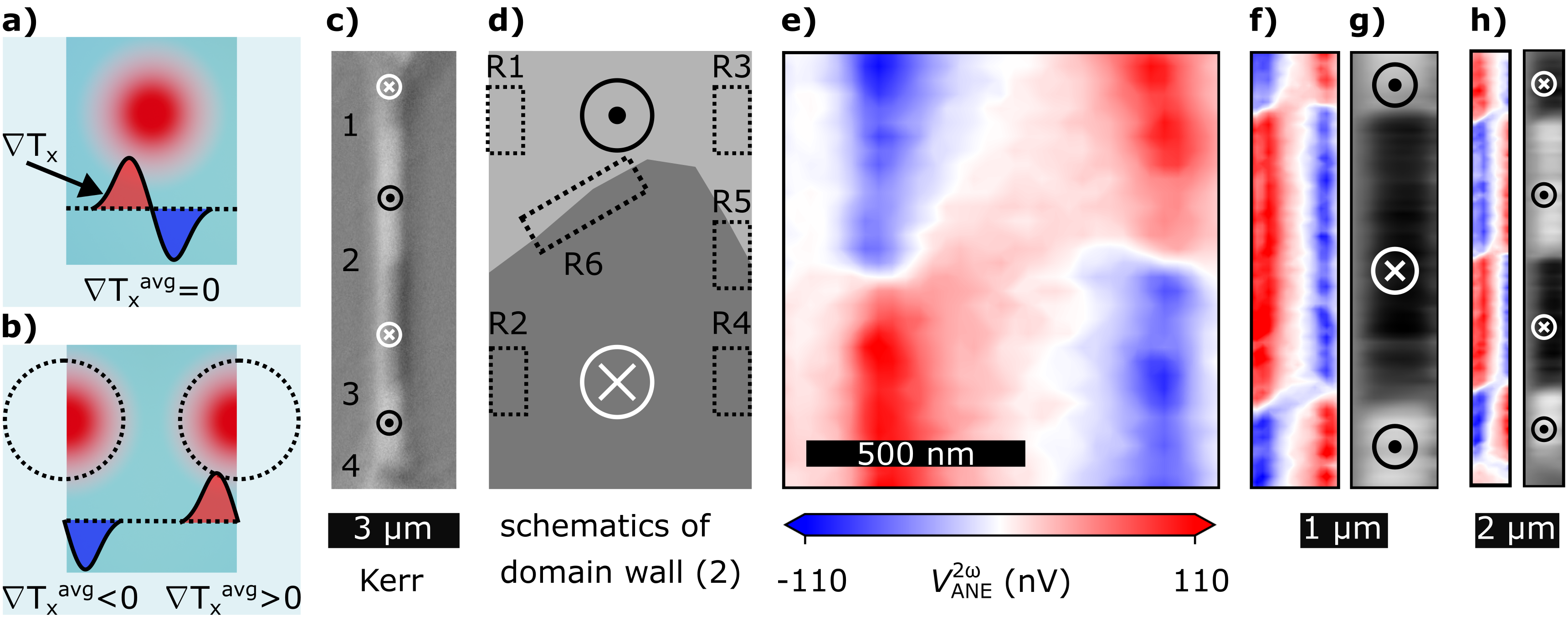}
\caption{(a) Schematic showing an OOP magnetic film on a transparent substrate. The laser illumination creates an asymmetric IP temperature gradient resulting in zero $\nabla T_{\mathrm{x}}^{\mathrm{avg}}$. (b) Schematic showing a laser beam focused near the edges of a metallic film that partially illuminates the film. In this case $\nabla T_{\mathrm{x}}^{\mathrm{avg}}$ is non zero and of opposite sign at the two edges. (c) Polar Kerr microscope image of the racetrack nanowire. The numbers indicate the different domain walls. (d) Schematics showing different regions around the magnetic domain wall (2)(e) Near-field-based ANE microscope image of the area surrounding the domain wall \#2. (f) Same as (e) for larger area consisting of domain walls \#2 and \#3. (g) ANE signal in f is integrated along the x-direction. (h) same as f,g for the ANE measurement performed across the full nanowire consisting of four domain walls.} 
\label{fig:nf_PMA_domain}
\end{figure}

At the edges of the sample, a non-zero $\nabla T_{\mathrm{x}}^{\mathrm{avg}}$ arises, as illustrated in \textbf{Fig.\ref{fig:nf_PMA_domain}(a,b)} Hence, one can expect that at the edges of an OOP magnetized sample a large ANE signal will be generated, while it should vanish in the center of the wire due to a vanishing $\nabla T_{\mathrm{x}}^{\mathrm{avg}}$. In addition, even for the case of zero $\nabla T_{\mathrm{x}}^{\mathrm{avg}}$ in the center region of the wire, a non uniform OOP magnetization (e.g. close to a domain wall) will result in a finite ANE signal, because the two lobes of the IP thermal gradient no longer cancel. This  can be seen from equation\eqref{eq2}.
We now demonstrate the validity of these conclusions by NF-SANE microscopy using a Co/Ni multilayer stack with perpendicular magnetic anisotropy \cite{PMA,PMA1}.
The Co/Ni stack is patterned into a \SI{700}{nm} wide nanowire. A multi-domain state is induced by using a decaying a.c. OOP magnetic field. The prepared sample consists of four domain walls within the narrow section of the wire, as confirmed by polar Kerr microscopy (\textbf{Fig.\ref{fig:nf_PMA_domain}(c)}). Subsequently, this magnetic structure is investigated with NF-SANE.

In agreement with our prediction of  $\nabla T_{\mathrm{x}}$, NF-SANE indeed shows large signals for the OOP magnetized sample that are related to the domain state. We now analyze this result in detail. We focus on the second domain wall in \textbf{Fig.\ref{fig:nf_PMA_domain}(c)} as schematically shown in \textbf{Fig.\ref{fig:nf_PMA_domain}(d)}. Here, different regions of interest are marked as R1-R6. R1 and R2 at the left edge are characterized by uniform $m_{\mathrm{z}}^{\mathrm{up}}$ and $m_{\mathrm{z}}^{\mathrm{down}}$. The negative $\nabla T_{\mathrm{x}}^{\mathrm{avg}}$ at this edge generates a negative $\mathrm{V_{ANE}}$ for $m_{\mathrm{z}}^{\mathrm{up}}$ in R1, and a positive $\mathrm{V_{ANE}}$ for $m_{\mathrm{z}}^{\mathrm{down}}$ in R2, as shown in \textbf{Fig.\ref{fig:nf_PMA_domain}(e)}. Similarly, a positive $\nabla T_{\mathrm{x}}^{\mathrm{avg}}$ at the right edge results in a positive $\mathrm{V_{ANE}}$ for $m_{\mathrm{z}}^{\mathrm{up}}$ in region R3 and a negative $\mathrm{V_{ANE}}$ for $m_{\mathrm{z}}^{\mathrm{down}}$ in R4. These findings demonstrate that the ANE near-field signal is sensitive to the OOP magnetization in the vicinity of the edges. 

Next, we consider region R5, which consist of a magnetic domain wall at the right edge of the wire. We perform a line scan in the $y$-direction. Since the ANE signal in this region is proportional to the magnetization, we can obtain the spatial resolution of the ANE microscope with such a line scan across the domain wall. We thereby find a spatial resolution of $\mathrm{66 \pm 10}$ nm (\textbf{S.9(b)}). Thus, we are able to employ NF-SANE to image domain walls in PMA magnetic structures with a resolution that is better than \SI{70}{nm}.

R6 consists of a domain wall in the middle of the wire. Illuminating this region creates a positive $\nabla T_{\mathrm{x}}^{\mathrm{avg}}$ on the left half of the laser spot. This generates a positive $\mathrm{V_{ANE}}$ with $m_{\mathrm{z}}^{\mathrm{up}}$. The right half of the laser spot creates a negative $\nabla T_{\mathrm{x}}^{\mathrm{avg}}$ that also generates a positive $\mathrm{V_{ANE}}$ with $m_{\mathrm{z}}^{\mathrm{down}}$. Thus, $m_z^{up}$ and $m_z^{down}$ on either side of the domain wall generate the same sign of voltage resulting in a net positive ANE signal, as we observe at the upper domain wall in \textbf{Fig.\ref{fig:nf_PMA_domain}(f)}. Similarly, a negative ANE voltage is observed at the lower domain wall because in this case the positive and negative $\mathrm{\nabla T_x}$ illuminates $m_{\mathrm{z}}^{\mathrm{down}}$ and $m_{\mathrm{z}}^{\mathrm{up}}$ domains, respectively. The vicinity of domain walls in PMA materials is characterized by a large lateral gradient of the OOP magnetization. This explains the observed domain wall contrast in the ANE signals.

Determining the magnetic structure in nanowires are perhaps the most important application of the NF-SANE imaging method. This specific geometry allows one to numerically integrate the observed ANE signal across the wire width to obtain a signal that is proportional to the magnetization and not only its gradient. The result of this procedure is shown in the gray scale images \textbf{Fig.\ref{fig:nf_PMA_domain}(g,h)} directly reflecting the domain structure in the wire.

Finally, we discuss the underlying mechanisms that make ANE microscopy advantageous when studying magnetic materials on the nanoscale. The magnitude of the ANE voltage in a given magnetic wire is proportional to $\frac{P}{w}$ (\textbf{Supplementary S.V}), where $P$ is the total absorbed power that contributes to heating the wire, and $w$ is the wire width. The magnitude of the ANE signal is independent of all other geometric factors. The laser-based heating utilized in ANE microscopy has two advantages in this regard. First, the entire absorbed energy is utilized to directly heat the wire, in contrast to resistive-heating methods where a majority of the heat energy is dissipated elsewhere \cite{ANE_res_heat_PR_Appl_2022,ANE_res_Ma2019,ANE_res_PhysRevB}. Second, since the voltage is inversely proportional to the wire width, studying narrower wires with ANE microscopy gives rise to larger signals. 

In addition, this result ($V_{\mathrm{ANE}}\propto\frac{P}{w}$) allows us to understand the magnitude of the nanoscale plasmonic enhancement of the optical field below the AFM tip \cite{SNOM_enhancement,SNOM_enhancement1}. For this, we compare the ANE signals obtained by means of SANE and NF-SANE using wires composed of the same material. We observe $V_{\mathrm{ANE}}$ = \SI{4}{\micro V} in a \SI{10}{\micro m} wide wire induced by a (\SI{5}{mW}$\times0.35=$\SI{1.75}{mW}) focused laser beam, taking the reflection coefficient of the CoFeB film to be 0.65 for the laser beam with a wavelength of 532 nm \cite{Co_reflt}. Based on this, our observation of a \SI{40}{nV} signal with the NF-SANE signal in a \SI{2}{\micro m} wide wire (see, for example, Fig. 3) would require a laser power of \SI{3.5e-3}{mW} underneath the AFM tip. In our NF-SANE setup, a 25 mW laser beam is focused by a 0.4 numerical aperture parabolic mirror. This illuminates the tip over a circular region of diameter \SI{10}{\micro m}, resulting in \SI{1.6e-3}{mW} at the \SI{80}{nm} wide laser spot underneath the tip. Considering a reflection coefficient of 0.95 for the given laser beam with a wavelength of \SI{8}{\micro m} \cite{Co_reflt}, the absorbed laser power amounts to only \SI{3.2e-4}{mW}. This means that the near-field interaction between the AFM tip and the sample enhances the laser power underneath the tip by a factor of approximately 44.

We have demonstrated the reliability of ANE microscopy to image magnetic domain states in in-plane magnetized wires down to the nanoscale. Our analysis also reveals the presence of a very large in-plane gradient that can be employed to probe out-of-plane magnetization and magnetic domain walls with a resolution of 70 nm, most suitable for racetrack nano wires. In addition, by using well-known magnetic textures, such as a vortex or a domain wall, one can analyze the spatial distribution of temperature gradients at the nanoscale.

{\bf Acknowledgements}

Financial support from the German research foundation (DFG) through collaborative research center (CRC) 227 is gratefully acknowledged. 

{\bf Author contributions}
The contribution from the authors is as follows: 

Conceptualization: GW, SSPP 

Methodology: AP, JD, JY, CK, RD, JMT 

Investigation: AP, JD, JY 

Visualization: AP

Project administration: GW, SSPP 
Supervision: GW 

Writing – original draft: AP, JMT, CK, RD, GW, SSPP 

Writing – review and editing: AP, JMT, CK, RD, GW, SSPP

{\bf Competing interests:}
The authors declare no competing financial or non-financial interests.


{\bf List of Supplementary Materials}
\textbf{Methods and sample fabrication:}
\begin{enumerate}
    \item[Sec.I.]Thin film growth and device fabrication 
    \item[Sec.II.] Details of the experimental setup for scanning ANE microscope. 
\end{enumerate} 

\textbf{Supplementary Figures:}

\begin{enumerate}
    \item [S2.] Dependence of the ANE signal magnitude on input laser power. 
    \item [S3.] Dependence of the ANE signal magnitude on width of the device wire structures.
    \item [S4.] Dependence of the ANE signal magnitude on the area of the input heat flux.
    \item [S5.] COMSOL simulations for out-of-plane heat gradient.
    \item [S6.] COMSOL simulations for in-plane heat gradient.
    \item [S7.] Intensity distribution of the focused laser beam.
    \item [S8.] Spatial distribution of the out-of-plane temperature gradient.
    \item [S9.] Spatial distribution of the in-plane temperature gradient.
    \item[S10.] Micro--nagnetic simulations. 
    \item [S11.] ANE based hysteresis measurement.

\end{enumerate}


%
\newpage
\section{Supplementary materials}
\renewcommand\thefigure{S\arabic{figure}}    
\setcounter{figure}{0}

\section{I. Sample fabrication}
The CoFeB films with thicknesses of \SI{15}{nm} and \SI{45}{nm} were deposited using an ultra-high vacuum magnetron sputtering system on (001)-cut MgO substrates. The films were capped with a \SI{2}{nm} Au protective layer. The base pressure of the sputtering system is \SI{3e-8}{mbar}, while the Ar process gas pressure during deposition was \SI{4.5e-3}{mbar}. The deposition rate was \SI{0.1}{A/s}. 

The sample with out-of-plane magnetization had the following layer structure: TaN (50)/Pt (12)/Co (3)/Ni (7)/Co (2)/Ni (7)/Co (2)/Ni (7)/Co (2)/TaN (30) A. These were deposited on sapphire substrates using a second ultra-high vacuum magnetron sputtering system with a base pressure of \SI{3e-9}{mbar}, while the Ar process gas pressure was \SI{3e-3}{mbar}.  Pt, Co, and Ni were deposited at rates of \SI{0.82}{A/s}, \SI{0.21}{A/s}, and \SI{0.22}{A/s}, respectively. All the deposition steps were performed at room temperature.

The wire structures for the CoFeB samples were defined using lithography and lift-off steps. Electron beam lithography was used for the devices with features smaller than \SI{2}{\micro m}, while the devices with larger structures were patterned using a mask-less optical lithography system. The out-of-plane magnetized Co/Ni stack was patterned using e-beam lithography and dry etching using Ar-ion bombardment.
\newpage
\section{II. Scanning ANE microscopy setup}

\begin{figure}[H]
\centering
\includegraphics[width=0.5\textwidth]{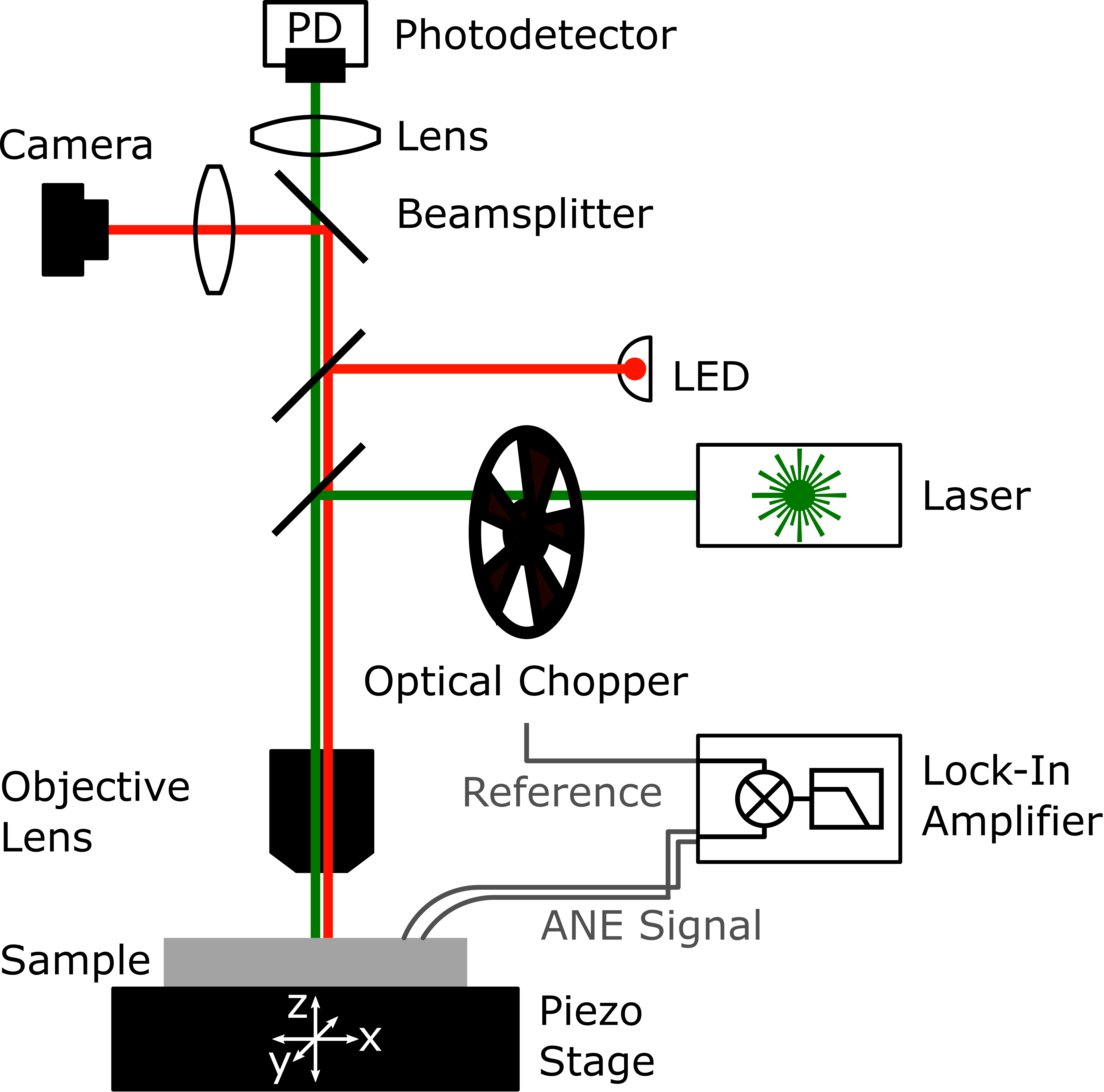}
\caption{The experimental setup for the scanning ANE (SANE) microscope consists of a \SI{532}{nm} cw laser beam from a diode-pumped solid state laser. The beam is focused on the sample by a high numerical aperture objective lens ($NA=0.7$). The intensity of the laser beam is modulated using an optical chopper. A piezo stage allows to scan the sample under the focal spot while the ANE signal is detected electrically. The intensity of the reflected laser beam can be measured with a photodetector.}
\label{fig:ff_Setup}
\end{figure}
\newpage

\section{III. Laser power dependence of the ANE signal}

\begin{figure}[H]
\centering
\includegraphics[width=1\textwidth]{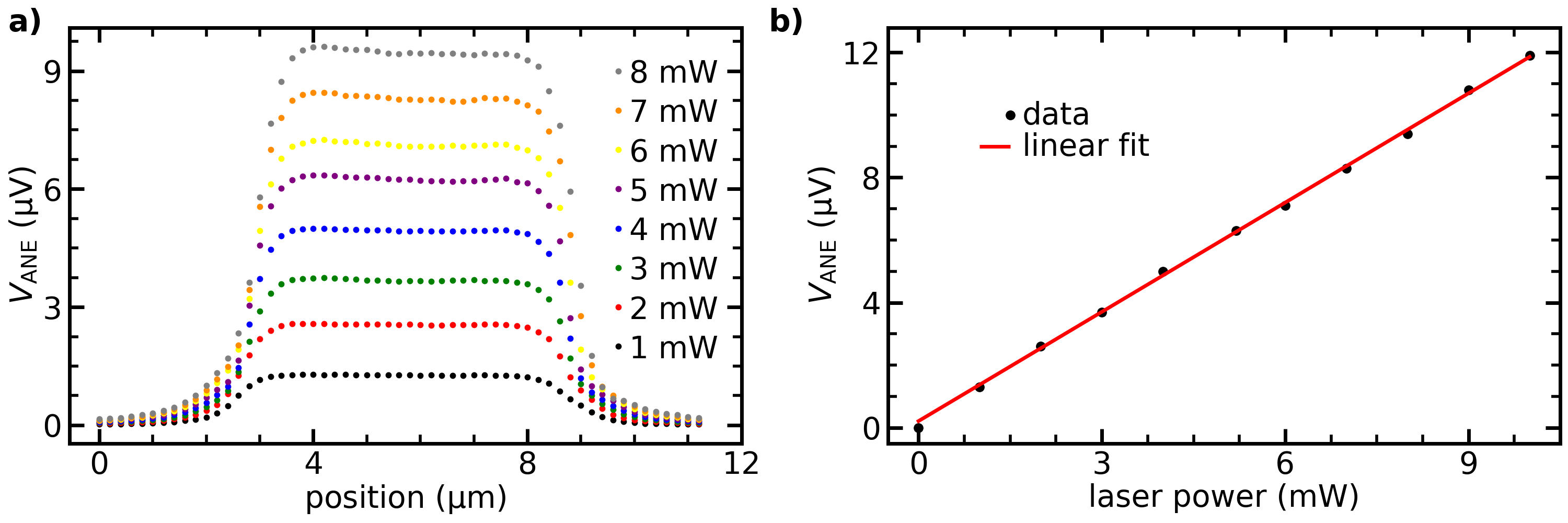}
\caption{(a) ANE line scan across a \SI{6}{\micro m} wide wire at different laser powers. (b) Average ANE signal across the wire width scales linearly with laser power.}
\label{fig:las_pow}
\end{figure}

\section{IV. Wire width dependence of the ANE signal}
\begin{figure}[H]
\centering
\includegraphics[width=1\textwidth]{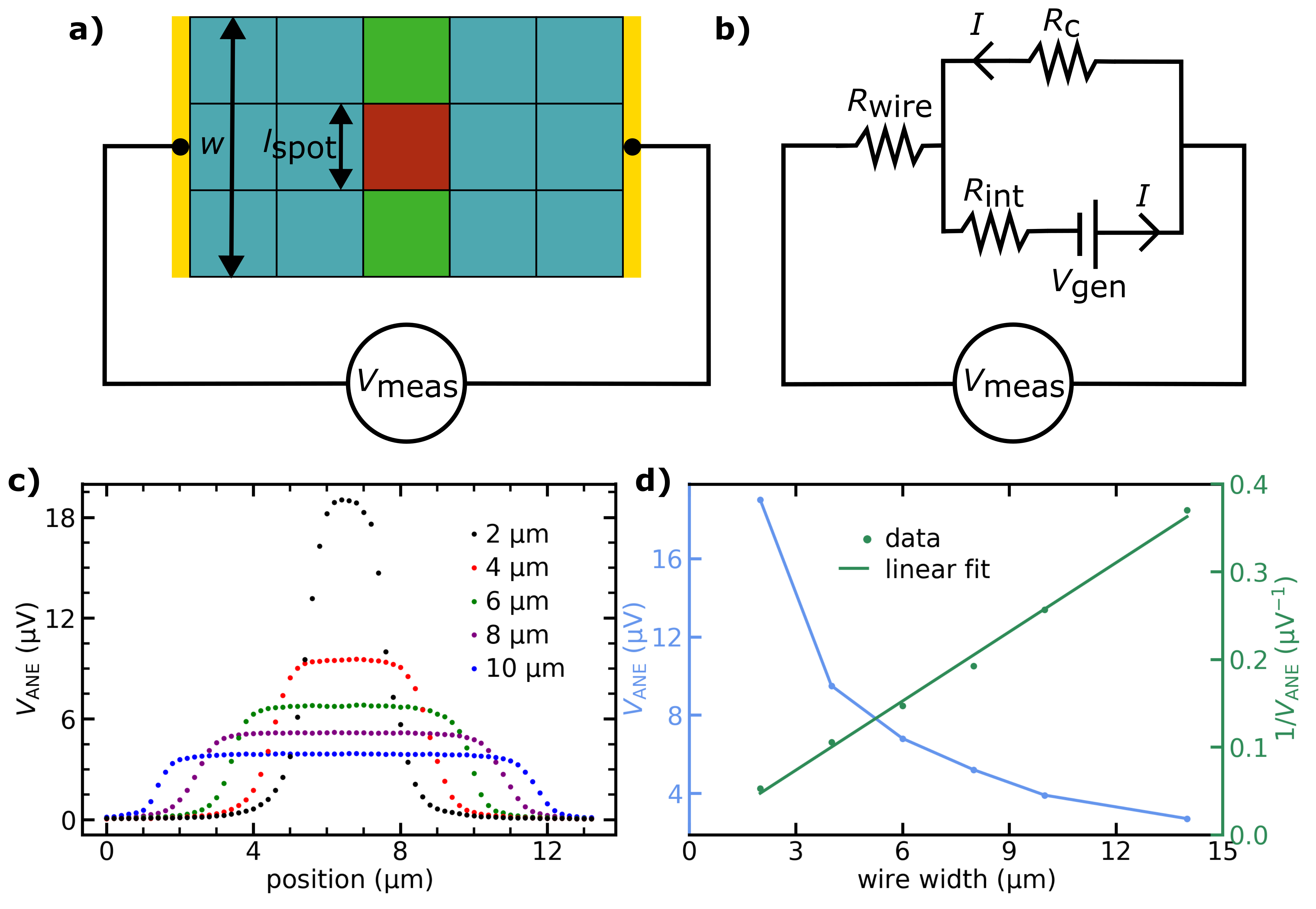}
\caption{(a) Schematic showing a magnetic wire of width $w$, with part of the wire across the dimension $l_{\mathrm{spot}}$ subjected to a heat flux (b) Equivalent electrical circuit diagram for the device in (a). (c) ANE line scans across magnetic wires of different widths. (d) Average $V_{\mathrm{ANE}}$ signal across the magnetic wires plotted against $w$ (blue, left axis) and $1/V_{\mathrm{ANE}}$ plotted against $w$ (green, right axis).}
\label{fig:wire_width}
\end{figure}

Having observed that the magnitude of the ANE signal depends on the wire width when other experimental parameters are kept identical, we developed an analytical model for this dependence and verify the model with the experimental data. We consider the scenario in which a part of the magnetic wire is subjected to a heat flux that creates an out-of-plane (OOP) temperature gradient. This is shown in red in \textbf{Fig.\ref{fig:wire_width}(a)}. The illuminated area (of dimension $l_{\mathrm{spot}}$) acts as a battery due to the ANE induced voltage ($V_{\mathrm{gen}}$). This battery has an internal resistance ($R_{\mathrm{int}}$) due to the finite resistance of the illuminated part of the wire. The regions of the wire shown in green provide a conducting path with resistance $R_{\mathrm{c}}$. The equivalent circuit diagram is shown in \textbf{Fig.\ref{fig:wire_width}(b)}, where $R_{\mathrm{wire}}$ indicates resistance of the remainder of the magnetic wire as well as the resistance of the connecting leads. 

We determine the influence of $R_{\mathrm{c}}$ and $R_{\mathrm{int}}$ on the magnitude of the measured ANE voltage ($V_{\mathrm{meas}}$). Assuming $R_{\mathrm{wire}}$ to be small (\SI{\approx1}{k\ohm}) relative to the lock-in impedance, its role can be ignored. Therefore $V_{\mathrm{meas}}$ is simply the potential across $R_{\mathrm{c}}$, which is equal to $I \cdot R_{\mathrm{c}}$, where $I$ is the current flowing in the circuit.
\newpage
\begin{align*}
V_{\mathrm{meas}} &= I\cdot R_{\mathrm{c}}\\
&=\frac{V_{\mathrm{gen}}}{R_{\mathrm{c}}+R_{\mathrm{int}}}\cdot R_{\mathrm{c}}  \\
&=\frac{R_{\mathrm{c}}}{R_{\mathrm{c}}+R_{\mathrm{int}}}\cdot V_{\mathrm{gen}} \\
\end{align*}

Since $R_{\mathrm{c}} \propto \frac{1}{w-l_{\mathrm{spot}}}$ and $R_{\mathrm{int}} \propto \frac{1}{l_{\mathrm{spot}}}$, where $w$ is the wire width,

\begin{align*}
V_{\mathrm{meas}}&=\frac{\frac{1}{w-l_{\mathrm{spot}}}}{\frac{1}{w-l_{\mathrm{spot}}}+\frac{1}{l_{\mathrm{spot}}}}\cdot V_{\mathrm{gen}} \\
&=\frac{l_{\mathrm{spot}}}{w}\cdot V_{\mathrm{gen}}
\end{align*}

Thus, we find that the measured ANE signal is inversely proportional to the wire width. ANE line scans across magnetic wires of different widths are shown in \textbf{Fig.\ref{fig:wire_width}(c)}. The inverse dependence of $V_{\mathrm{ANE}}$ on $w$ is demonstrated in \textbf{Fig.\ref{fig:wire_width}(d)} and verified by a linear fit of $1/V_{\mathrm{ANE}}$ against $w$.  

\newpage

\newpage
\section{V. Dependence of ANE signal on heat flux area}
\begin{figure}[H]
\centering
\includegraphics[width=1\textwidth]{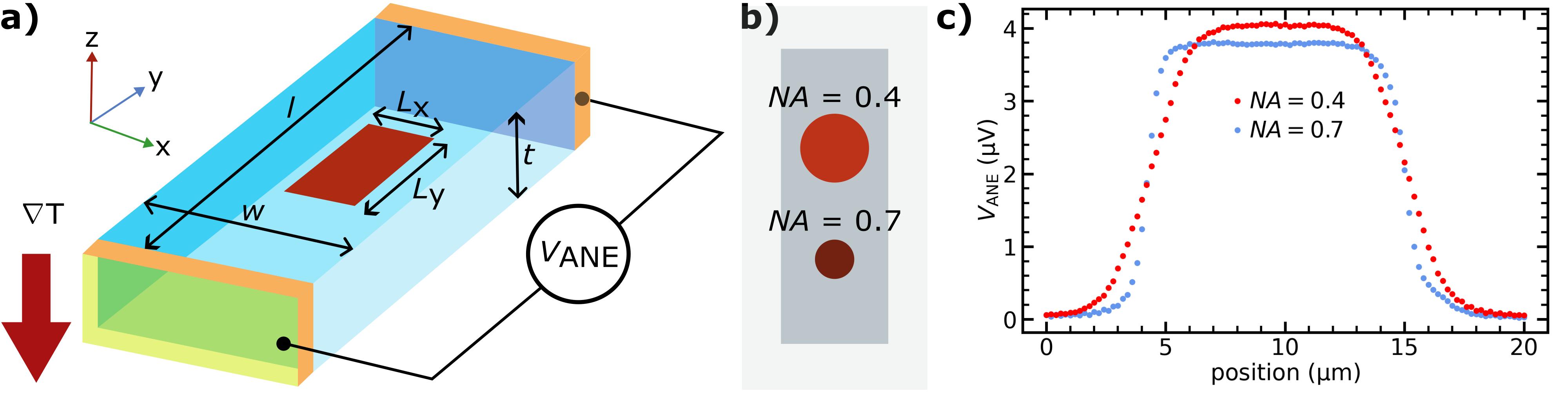}
\caption{(a) Schematic showing a magnetic wire subjected to a heat flux in a square area (red). (b) Focal spot created by a focused laser beam. The heat flux area created by the $NA=0.4$ objective is larger than that created by the $NA=0.7$ objective. (c) ANE line scans across a magnetic wire using these two different objective lenses.}
\label{fig:NA_dep}
\end{figure}

We now consider a magnetic wire of width $w$ and thickness $t$ subjected to a heat flux of $P$ in Watts across a $L_{\mathrm{x}} \cdot L_{\mathrm{y}}$ rectangular area (\textbf{Fig.\ref{fig:NA_dep}(a)}). We determine if $L_{\mathrm{x}}$ or $L_{\mathrm{y}}$ influences the magnitude of the measured ANE signal $V_{\mathrm{ANE}}$. The heat flux creates an out-of-plane temperature gradient $\nabla T_{\mathrm{z}}$. This results in an ANE signal generated along the $x$-direction ($\mathrm{V_{gen}^x}$). 
In terms of the ANE coefficient $S_\mathrm{{ANE}}$, it can be written as $V_\mathrm{{gen}}^{\mathrm{x}}$ =  $\mu \cdot S_\mathrm{{ANE}}\cdot \nabla T_{\mathrm{z}}\cdot L_{\mathrm{y}}$. The temperature gradient is proportional to the heat flux density, giving,  

\begin{align*}
\mathrm{V_{gen}^x} &=  \mu\cdot S_\mathrm{{ANE}}\cdot \nabla T_{\mathrm{z}}\cdot L_{\mathrm{y}}\\
&\propto \mu\cdot S_\mathrm{{ANE}}\cdot\frac{P}{L_{\mathrm{x}}L_{\mathrm{y}}} \cdot L_{\mathrm{y}} \\
&\propto \frac{\mu \cdot S_\mathrm{{ANE}}\cdot P}{L_{\mathrm{x}}} 
\end{align*}

$V_\mathrm{{ANE}}$ is the measured ANE voltage along $x$-direction ($V_{\mathrm{meas}}^{\mathrm{x}}$), that is related to $V_\mathrm{{gen}}^{\mathrm{x}}$ by

\begin{align*}
V_\mathrm{{ANE}} &=V_{\mathrm{meas}}^{\mathrm{x}}\\
&=\frac{L_{\mathrm{x}}}{w}\cdot V_\mathrm{{gen}}^{\mathrm{x}}\\
&\propto \frac{L_{\mathrm{x}}}{w}\cdot \frac{\mu \cdot S_\mathrm{{ANE}}\cdot P}{L_{\mathrm{x}}}\\
&\propto \frac{\mu \cdot S_\mathrm{{ANE}} \cdot P}{w}
\end{align*}

Thus, we see that the magnitude of the ANE signal only depends on the total incident power and the wire width, and is independent of the heat flux area. To verify this experimentally, we heat the magnetic wire with constant power over areas of different dimensions. This is achieved by focusing a laser beam using objective lenses with two different numerical apertures ($NA$) of 0.4 and 0.7. Since the diffraction limited spot size of the focused laser beam is inversely proportional to the $NA$, these yield different heat flux areas, as schematically depicted in \textbf{Fig.\ref{fig:NA_dep}(b)}. Although, the heat flux area is larger for $NA=0.4$ relative to $NA=0.7$, ANE line scans across the magnetic wire show that the magnitude of ANE signal is identical for $NA=0.7$ and $NA=0.4$ (\textbf{Fig.\ref{fig:NA_dep}(c)}), confirming our analytical model.
\newpage

\section{VI. Laser heating induced temperature gradient}

\begin{figure}[H]
\centering
\includegraphics[width=1\textwidth]{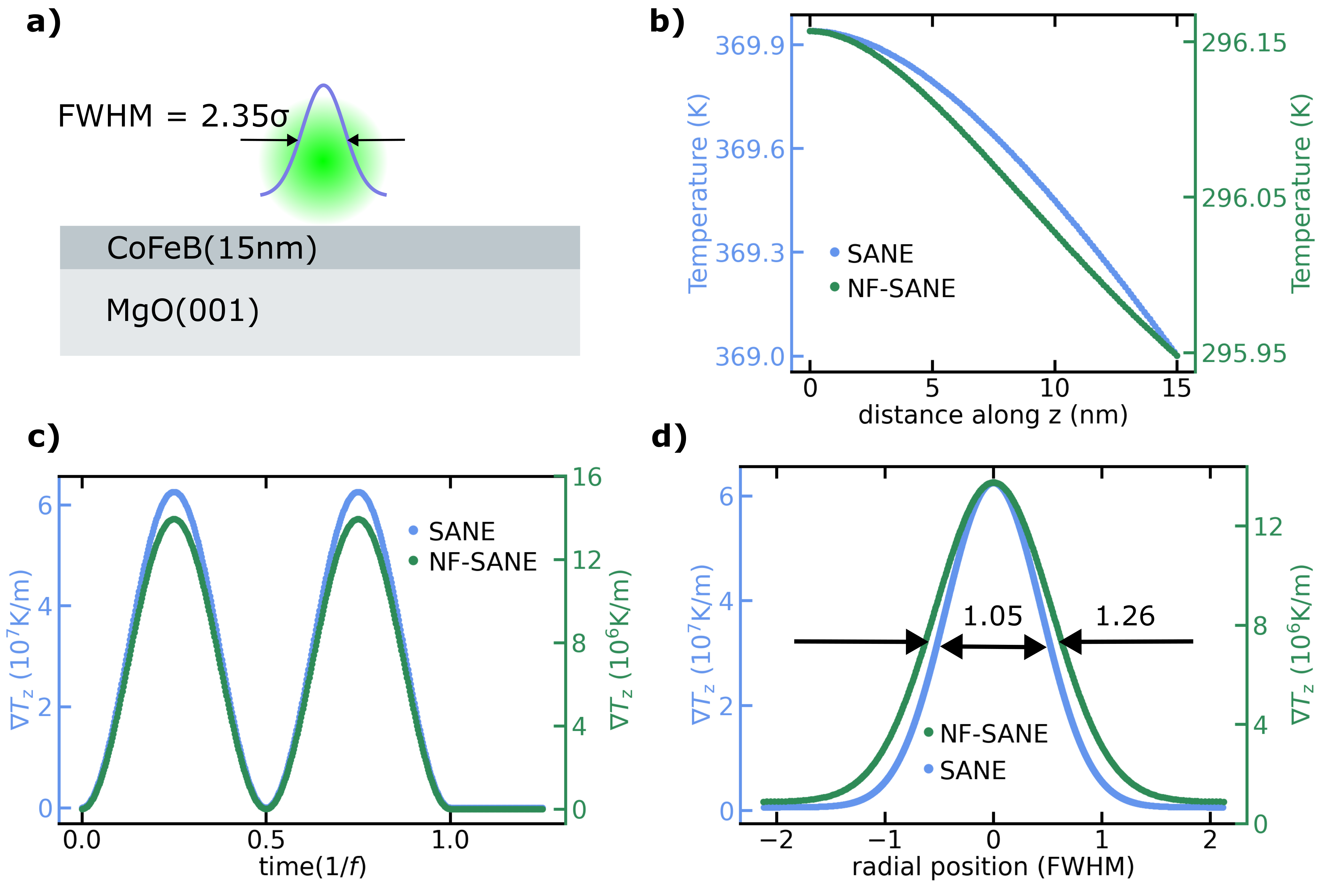}
\caption{(a) Schematic showing a metallic film on a transparent substrate illuminated by a Gaussian laser beam. (b) Line graph of temperature across the film thickness at time = $T$/4. (c)  (d) Time dependence of the average temperature gradient ($\nabla T_{\mathrm{z}}$) across the film. (d) Radial distribution of $\nabla T_{\mathrm{z}}$. The length scale on the horizontal axis is normalised to the FWHM of the laser intensity distribution. In b-d, left (blue) axis shows the simulation results for the SANE scenario, right (green) axis for the NF-SANE case.} 
\label{fig:heat_grad}
\end{figure}

We use finite element modelling simulations in COMSOL multi physics to estimate the spatial and temporal temperature distribution in a \SI{15}{nm} CoFeB film subjected to laser induced heating(\textbf{Fig.\ref{fig:heat_grad}(a)}). The beam from a \SI{532}{nm} cw laser is focused by an objective lens with $NA = 0.7$. For a diffraction limited spot in SANE, the spatial distribution of the laser intensity $I(r)$ can be modelled as a Gaussian beam, with FWHM ($2.33\mathrm{\sigma}$) = $\frac{\lambda}{2NA}$ . The laser intensity in near-field case (NF-SANE) is modelled as a Gaussian beam with FWHM approximated to be the AFM tip radius $=\SI{20}{nm}$.   

For an input laser power of $P$ (in mW) on the sample, the intensity distribution is,
\begin{equation*}
    I(r)= \frac{P}{2\pi\cdot \sigma^2}\cdot e^{-\left(\frac{r^2}{2\sigma^2}\right)} 
           \qquad \mathrm{[W/m^2}. 
\end{equation*}

Taking the absorption coefficient of our CoFeB films to be $\alpha$, a fraction of the beam ($\alpha P$) passes in the $z$-direction through the film and is attenuated exponentially. Thus we have 
\begin{equation*}
    I(r,z) = \frac{\alpha\cdot P}{2\pi\cdot\sigma^2}\cdot e^{-\left(\frac{r^2}{2\sigma^2}+\frac{z}{z_0}\right)}\qquad\mathrm{[W/m^2},
\end{equation*}
where $z_{\mathrm{0}}$ = \SI{10}{nm} is the skin depth of the CoFeB film.  The laser beam intensity absorbed within a thickness $\delta z$ is $\frac{dI}{dz}\cdot\delta z$. Thus, the absorbed laser beam energy per unit volume is $\frac{dI}{dz}$, giving:
\begin{equation*}
    Q(r,z) = \frac{dI}{dz}=\frac{\alpha\cdot P}{2\pi\cdot\sigma^2z_0}\cdot e^{-\left(\frac{r^2}{2\sigma^2}+\frac{z}{z_0}\right)}\qquad\mathrm{[W/m^3]}.
\end{equation*}
This acts as a local heat source, which is modulated at the frequency $f = 1/T$ and can be modelled as $sin^2\left(\frac{2\pi t}{T}\right)$. The spatial and temporal dependence of the heat source can therefore be written as:
\begin{equation*}
    Q(r,z,t) = \frac{\alpha\cdot P}{2\pi\cdot\sigma^2 \cdot z_0} \cdot e^{-\left(\frac{r^2}{2\sigma^2}+\frac{z}{z_0}\right)} \cdot sin^2\left(\frac{2\pi t}{T}\right)\qquad\mathrm{[W/m^3]}
\end{equation*}

The CoFeB film temperature rises due to this heat source and subsequently the absorbed heat diffuses through to the MgO substrate. We use the heat conduction module in COMSOL to solve this diffusion equation numerically. We simulated two different scenarios, representing the SANE ($\alpha \cdot P=$ \SI{4.5}{mW}, $\mathrm{\sigma} =$ \SI{190}{nm}, $f=$ \SI{600}{Hz}) and NF-SANE ($\alpha\cdot P=$ \SI{e-2}{mW}, $\mathrm{\sigma}=$ \SI{10}{nm}, $f=$ \SI{580}{kHz}) cases. In the SANE scenario, the temperature of the film directly underneath the laser spot rises to 370 K at time = $T/4$, when the laser power is at a maximum. A line graph along the $z$ direction shows a monotonous decrease of the temperature, resulting in a net gradient of approximately \SI{1}{K} across the \SI{15}{nm} film \textbf{Fig.\ref{fig:heat_grad}(b)}. Similar results are obtained in the NF-SANE case as well. 

The measured ANE signal is proportional to the average temperature gradient across the film $\nabla T_{\mathrm{z}} = \frac{T_{\mathrm{top}}-T_{\mathrm{bottom}}}{t}$. To investigate the temporal distribution of the thermal gradient, $\nabla T_{\mathrm{z}}$ is plotted against time. It follows the laser intensity modulation in both cases. (\textbf{Fig.\ref{fig:heat_grad}(c)}). This demonstrates that we can generate a modulated temperature gradient at both, the frequency of the optical chopper (used in the SANE microscope) and the vibrating AFM tip (used in the NF-SANE microscope). 
\newpage

\begin{figure}[H]
\centering
\includegraphics[width=1\textwidth]{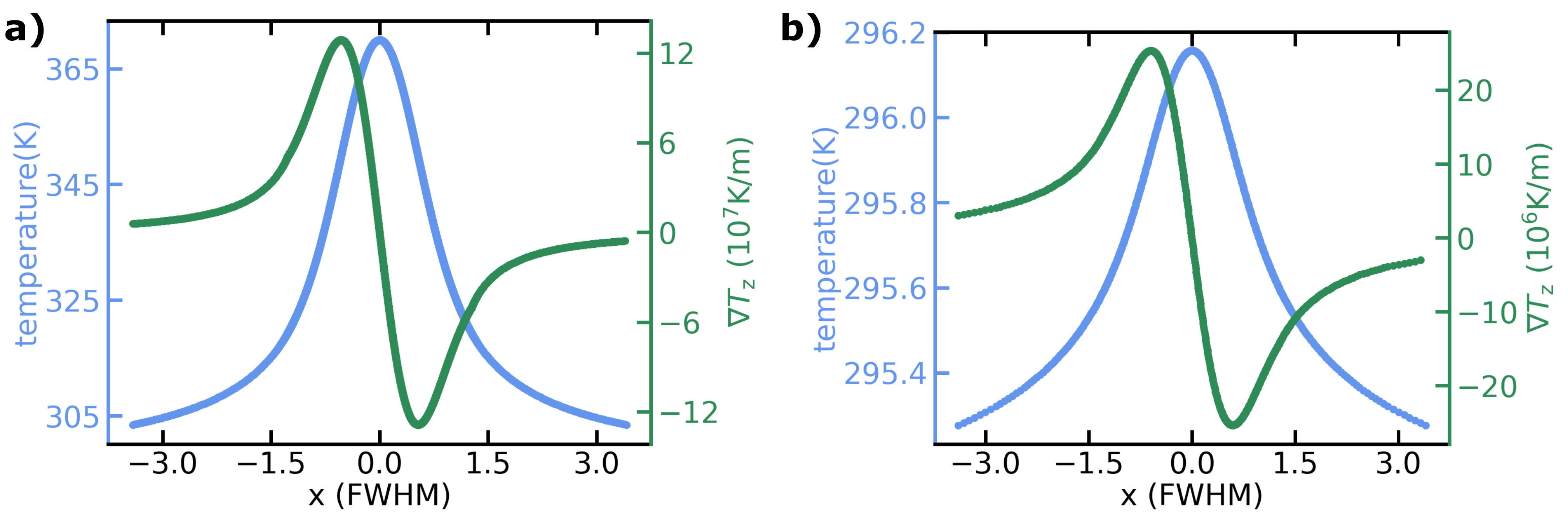}
\caption{(a) Temperature underneath the laser spot in the SANE microscope setup at time = $T/4$: line profile of the film temperature along the x-direction (blue, left axis), and IP temperature gradient $\nabla T_{\mathrm{x}}$ (green, right axis)(b) Same for the NF-SANE case. The length scale on the horizontal axis is normalised to the FWHM of the laser intensity distribution.}
\label{fig:heat_grad1}
\end{figure}

We also study the spatial distribution of $\nabla T_{\mathrm{z}}$ across the plane of the film. The radial distribution of $\nabla T_{\mathrm{z}}$ underneath the laser spot is fitted to a Gaussian function and the fit is plotted in (\textbf{Fig.\ref{fig:heat_grad1}(d)}). We find that in the far-field case, FWHM of $\nabla T_{\mathrm{z}}$ is only 5\% higher relative to FWHM of the laser intensity distribution. Similarly, for the near-field situation, the gradient distribution spread is only 26\% higher relative to laser intensity distribution. Thus a $\nabla T_{\mathrm{z}}$ arises only in the area directly underneath the laser spot. 

The spatial distribution of the film temperature in the $x$ direction is shown in (\textbf{Fig.\ref{fig:heat_grad1}}). The derivative of this curve gives the in-plane (IP) temperature gradient along the $x$-direction ($\nabla T_{\mathrm{x}}$). We find that this gradient is antisymmetric and that the net $\nabla T_{\mathrm{x}}$ across the laser beam profile is zero. However, when only part of the laser beam is considered, the net $\nabla T_{\mathrm{x}}$ can be non-zero and comparable in magnitude to that of the out-of-plane temperature gradient.

\section{VII. Intensity distribution of the focused laser beam}
\begin{figure}[H]
\centering
\includegraphics[width=1\textwidth]{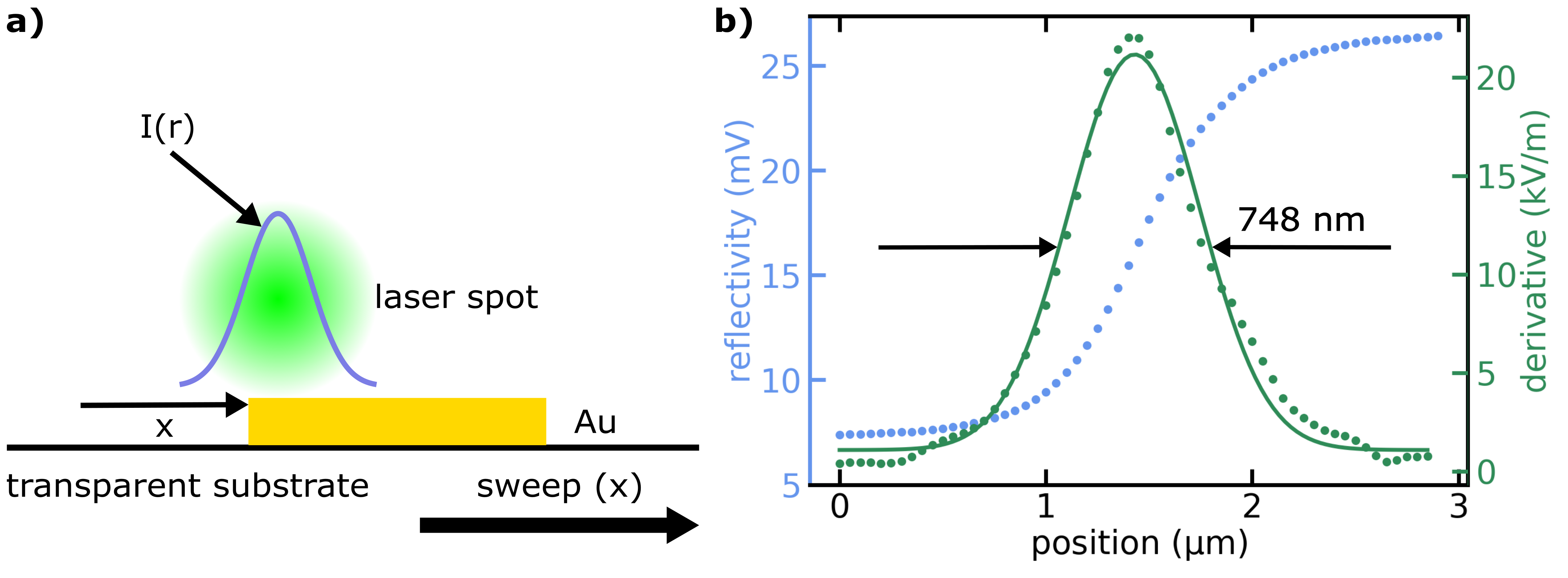}
\caption{(a) Schematics showing a focused laser beam swept across a gold film on a transparent substrate. (b) Plot of reflectivity against the laser spot position (blue, left axis). Spatial derivative of the reflectivity together with the Gaussian fit is shown (green, right axis).}
\label{fig:ff_spot_size}
\end{figure}

We determine the intensity distribution ($I(r)$) of the focused laser beam utilised in SANE microscope. This is obtained by scanning over one of the edges of a thin gold film with the laser spot and measuring the intensity of the reflected beam ($I_\mathrm{ref}$). $I_\mathrm{ref}$ as a function of edge position ($x$) can be written as 
\begin{equation*}
   I_\mathrm{ref}(x)=\int_{-\infty}^xr_1I(r)dr+\int_x^{\infty}r_2I(r)dr
\end{equation*}
Where, $r_1$ and $r_2$ are the reflection coefficient of the transparent substrate and the gold film respectively. Defining $\int I(r)dr= I^{int}(r)$, we have: 
\begin{equation*}
    I_\mathrm{ref}(x)=(r_1-r_2)\cdot I^{int}(x)+constant
\end{equation*}
Taking the derivative of above equation give, 
\begin{equation*}
    \frac{\mathrm{d}Ref(x)}{\mathrm{d}x}=(r_1-r_2)\cdot \frac{\mathrm{d}I^{int}(x)}{\mathrm{d}x}=(r_1-r_2)\cdot I(x)
\end{equation*}
Thus we see that laser intensity distribution $I(x)$ is proportional to derivative of the reflectivity line scan. \textbf{Fig.\ref{fig:ff_spot_size}(b)} shows reflectivity measurement as the edge is scanned across the focused laser beam spot. The derivative of this line scan indicates the laser intensity distribution, which is well described by a Gaussian distribution with FWHM = $748\pm22$ nm.  

\section{VIII. Generalised expression for ANE Signal in ANE microscope}

Consider a magnetic slab of the dimension $l\times w\times t$ (\textbf{Fig.\ref{fig:NA_dep}(a)}) illuminated by a laser beam. A temperature gradient $\nabla T_{\mathrm{x}}$ and $\nabla T_{\mathrm{z}}$ induces an electrical response along the slab length ($y$-direction) due to ANE:
\begin{equation*}
    E_{\mathrm{ANE}} = \mu \cdot S_{\mathrm{ANE}} \cdot\left(m_{\mathrm{x}}\nabla T_{\mathrm{z}} + m_{\mathrm{z}}\nabla T_{\mathrm{x}}\right).
\end{equation*}
Since the gradient is non-uniform under the laser spot, we divide the entire region into cubes of dimension $dx \times dy \times dz$ in which the gradient can be assumed to be constant. The resulting voltage is 
\begin{equation*}
    V_{\mathrm{ANE}} = E_{\mathrm{ANE}}\,dy = \mu \cdot S_{\mathrm{ANE}} \cdot\left(m_{\mathrm{x}}\nabla T_{\mathrm{z}} + m_{\mathrm{z}}\nabla T_{\mathrm{x}}\right) \, dy.
\end{equation*}
These stacks of the cubes act as batteries conducting in parallel along $x$ and $z$ direction and in series along $y$ direction. The net voltage averages out in $x$ and $z$ directions and adds up in $y$-direction. 
\begin{equation*}
    V_{\mathrm{ANE}} = \frac{\mu \cdot S_{\mathrm{ANE}}}{w\cdot t}\cdot \left( \int_0^w\int_0^l\int_o^tm_{\mathrm{x}}\nabla T_{\mathrm{z}}\,dxdydz + \int_0^w\int_0^l\int_o^tm_{\mathrm{z}}\nabla T_{\mathrm{x}}\,dxdydz\right).  
\end{equation*}
Assuming $m_{\mathrm{x}}$ and $m_{\mathrm{z}}$ to be uniform across the film thickness of a few nanometers, we have 
\begin{equation*}
    \int_0^tm_{\mathrm{x}}\nabla T_\mathrm{z}\,dz = m_{\mathrm{x}}\int_0^t\nabla T_\mathrm{z}\,dz = m_{\mathrm{x}}\cdot (T_\mathrm{top}-T_\mathrm{bottom}).
\end{equation*}
The temperature across the film is nearly identical, for example 369 K and 370 K in (\textbf{Fig.\ref{fig:heat_grad}(b)}). $\nabla T_{\mathrm{x}}$ can be assumed to be nearly constant across the film thickness, giving $\int_0^tm_{\mathrm{z}}\nabla T_\mathrm{x}\, dz = m_{\mathrm{z}}\nabla T_\mathrm{x}t$. Therefore, 
\begin{equation*}
    V_{\mathrm{ANE}} = \frac{\mu\cdot S_{\mathrm{ANE}}}{w}\cdot \left( \frac{1}{t}\cdot \int_0^w\int_0^lm_{\mathrm{x}}\cdot(T_\mathrm{top}-T_\mathrm{bottom})\, dxdy + \int_0^w\int_0^lm_{\mathrm{z}}\nabla T_{\mathrm{x}}\,dxdy\right) 
\end{equation*}

\section{IX. Spatial distribution of the temperature gradient}
\begin{figure}[H]
\centering
\includegraphics[width=1\textwidth]{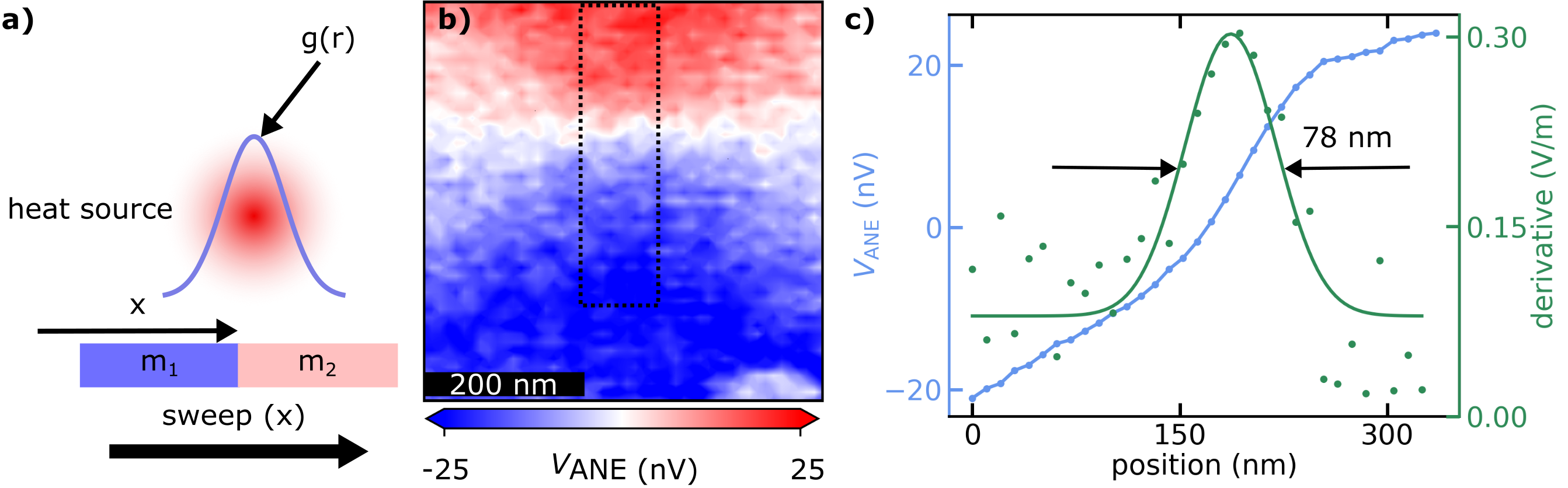}
\caption{(a) Schematics showing a magnetic domain wall across a heat source.(b) Repetition of the measurements in the main text \textbf{Fig.4 (c)}. (b) ANE line scan in the region shown by the dotted rectangle in (a), the right, green axis shows line scan derivative and the fit to a Gaussian.}
\label{fig:nf_resln}
\end{figure}

Similar to laser intensity distribution, we can determine temperature gradient distribution $g(r)$ by scanning a domain wall across the heat source spot (\textbf{Fig.\ref{fig:nf_resln}}(a)). Dependence of resulting ANE voltage $V_{\mathrm{ANE}}(x)$ on a domain wall position ($x$) is measured. In this case, the gradient distribution $g(x)$ would be proportional to the ANE derivative $\frac{\mathrm{d}V_{\mathrm{ANE}}(x)}{\mathrm{d}x}$. \textbf{Fig.\ref{fig:nf_resln}}(b) shows the repetition of the measurement shown in the main text \textbf{Fig.4(c)}. A line scan across the the vortex is shown in \textbf{Fig.\ref{fig:nf_resln}}(b), it's derivative indicates a temperature gradient distribution $g(x)$, which is fitted to a Guassian function with FWHM = \SI{78}{nm} 

\section{X. Determination of the spatial resolution for the AFM based ANE microscope with domain wall in PMA wire}
\begin{figure}[H]
\centering
\includegraphics[width=1\textwidth]{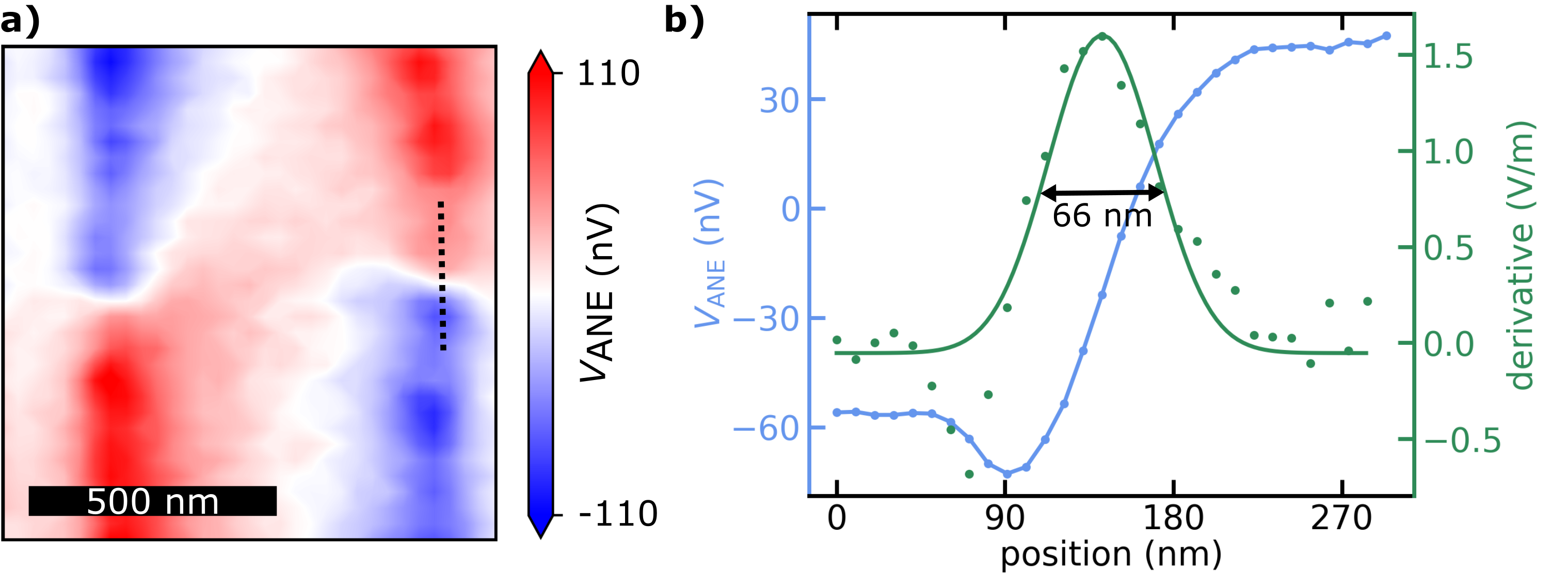}
\caption{(a) \textbf{Fig. 5(d)} in the main text.(b) ANE line scan across a magnetic domain wall shown by the dotted line in (a). The numerical derivative of the line scan is fitted to a Gaussian function.}
\label{fig:PMA}
\end{figure}

\section{XI. Micromagnetic simulations}
For the micromagnetic simulation using MuMax3, CoFeB magnetic structures were discretized in cells of size \SI{512}{nm^3}, thus the step size (\SI{8}{nm}) is less than the exchange length ($\lambda_{exc}=$\SI{8.6}{nm}) of $\mathrm{Co_{20}Fe_{60}B_{20}}$. We have taken $M_{\mathrm{S}} =$ \SI{65}{kAm^{-1}}, in-plane magnetic anisotropy constant $K =$ \SI{2.02}{kJm^{-3}}, exchange constant $A=$\SI{20}{pJm^{-1}} , damping coefficient $\alpha = 0.03$. We initialed with random magnetization and then applied the exponentially decreasing oscillating magnetic field to create a vortex ground state. 
The Runge-Kutta (RK45) method was used for the relaxation process.
\newpage
\begin{figure}[H]
\centering
\includegraphics[width=0.5\textwidth]{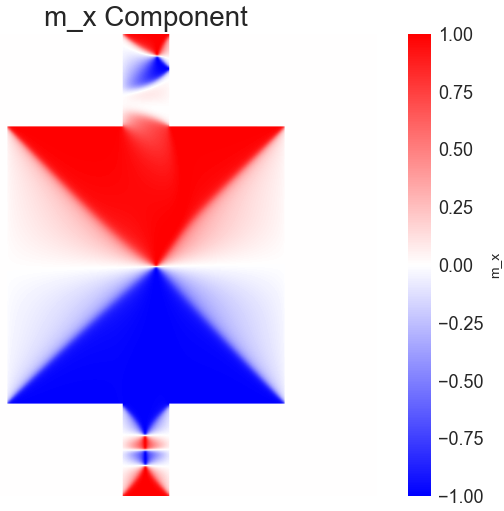}
\caption{Micromagnetic simulation for the the device discussed in the main text figure.4.}
\label{fig:mm_sim}
\end{figure}

\section{XII. Magnetic characterization with SANE microscopy}

\begin{figure}[H]
\centering
\includegraphics[width=0.5\textwidth]{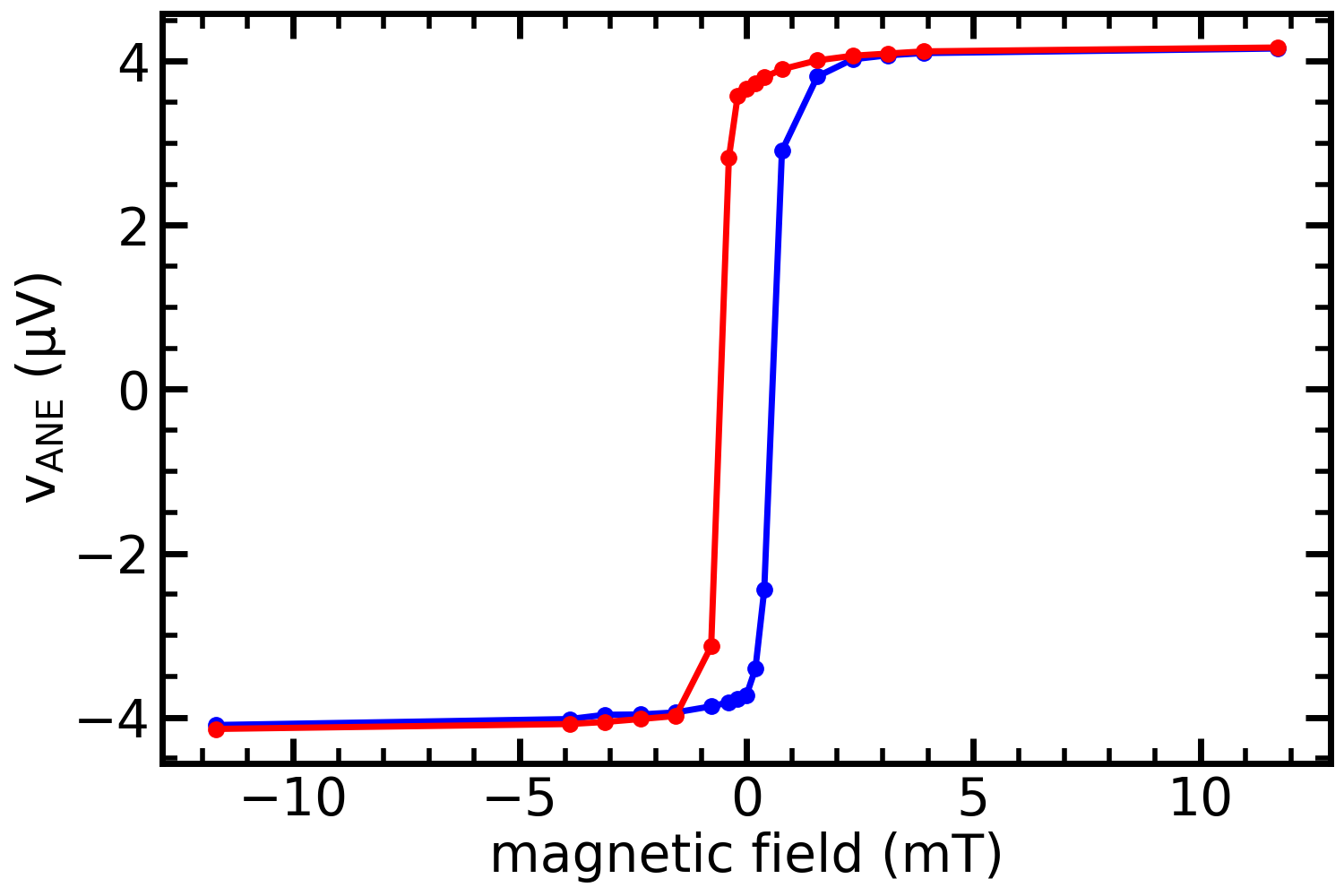}
\caption{Hysteresis measurement of the 15 nm thick CoFeB film with SANE microscopy}
\label{fig:hyst}
\end{figure}

We  demonstrate that the SANE microscope can be utilised for characterizing magnetic properties, for example a hysteresis loop measurement. We do so by scanning the laser over a  \SI{15}{\micro m}$\times$\SI{5}{\micro m} area and averaging $V_{\mathrm{ANE}}$ at different applied field strengths, as shown in \textbf{Fig.\ref{fig:hyst}}. For this measurement, a 15 nm thick CoFeB film with width $w =$\SI{14}{\micro m} and length $l=$ \SI{25}{\micro m} is chosen. The wider wire reduces shape anisotropy and provides reasonable coercive field to observe a hysteresis. 

\section{XIII. Magnetic property of the Co/Ni film}
Here we show hysteresis measurement for the Co/Ni multi stack PMA film. The device made out of this film was utilized to show ANE domain imaging in out-of-plane magnetized sample. The measurements are performed with a vibrating-sample magnetometer.   

\begin{figure}[H]
\centering
\includegraphics[width=0.5\textwidth]{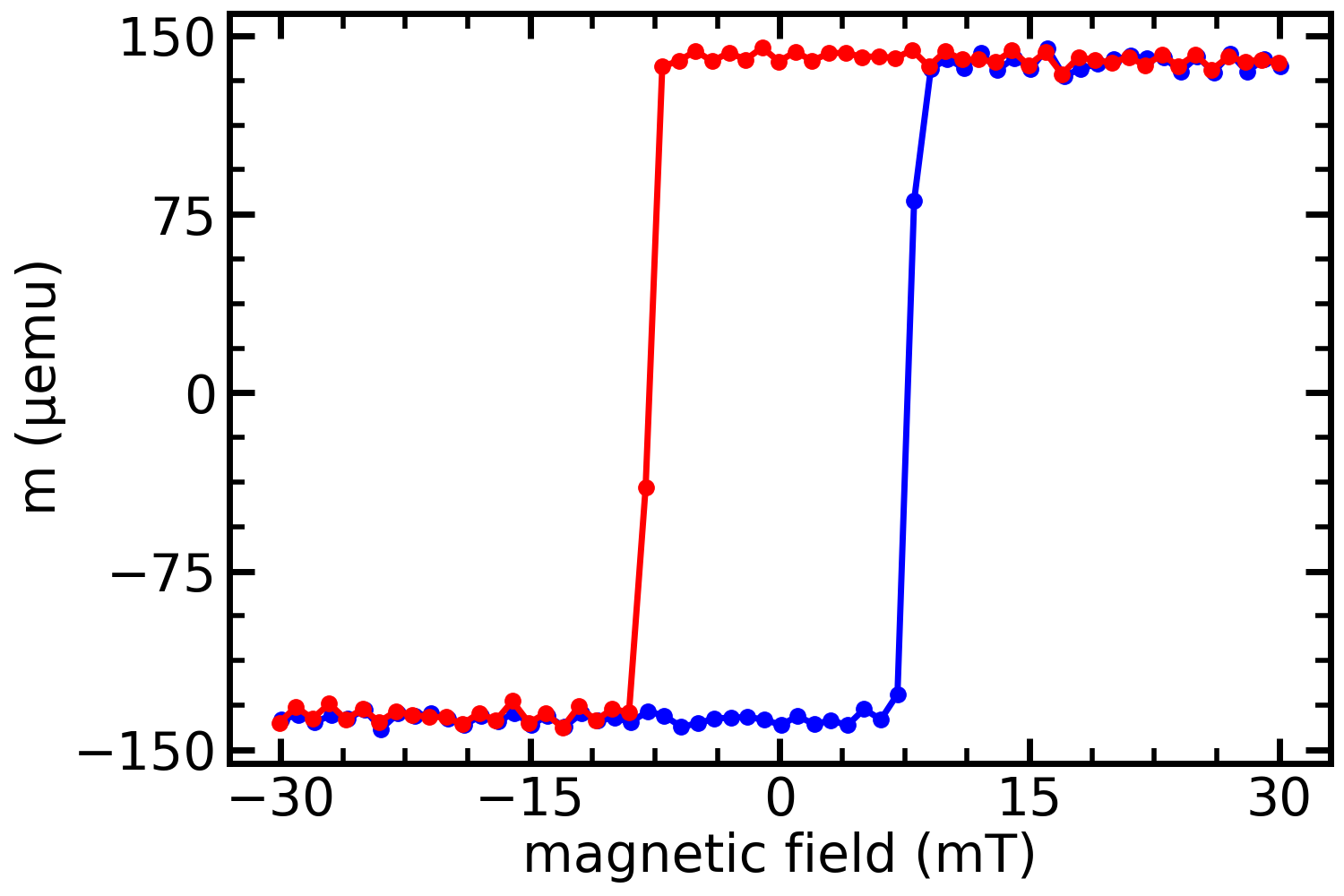}
\caption{Hysteresis measurement of the Co/Ni film.}
\label{fig:hyst_pma}
\end{figure}
 

\end{document}